\documentstyle{cupconf}
\input psfig.sty

\ifoldfss
\else
  \ifnfssone
    \newmathalphabet{\mathit}
      \addtoversion{normal}{\mathit}{cmr}{m}{it}
      \addtoversion{bold}{\mathit}{cmr}{bx}{it}
    \newmathalphabet{\mathcal}
      \addtoversion{normal}{\mathcal}{cmsy}{m}{n}
    \else
    \ifnfsstwo
    \fi
  \fi
\fi

\def\hexnumber#1{\ifcase#1 0\or1\or2\or3\or4\or5\or6\or7\or8\or9\or
 A\or B\or C\or D\or E\or F\fi }

\makeatletter
\ifx\CUP@mtlplain@loaded\undefined
\else
\fi
\makeatother
 \makeatletter
 \ifx\CUP@mtlplain@loaded\undefined
   \font\tenbmi=cmmib10 at 10pt
   \font\sevenbmi=cmmib10 at 7pt
   \font\fivebmi=cmmib10 at 5pt

   \newfam\bmifam
   \textfont\bmifam=\tenbmi
   \scriptfont\bmifam=\sevenbmi
   \scriptscriptfont\bmifam=\fivebmi
   
 \fi
 \makeatother
\ifnfsstwo

\fi
\ifnfssone

\fi
\ifoldfss

\fi

\mathchardef\varLambda="0103

\makeatletter
\ifx\CUP@mtlplain@loaded\undefined
\else
\fi
\makeatother
\makeatletter
\ifx\CUP@mtlplain@loaded\undefined
  \font\tenbms=cmbsy10
  \font\sevenbms=cmbsy10 at 7pt
  \font\fivebms=cmbsy10 at 5pt
  \newfam\bmsfam
  \textfont\bmsfam=\tenbms
  \scriptfont\bmsfam=\sevenbms
  \scriptscriptfont\bmsfam=\fivebms

  \edef\bsy@{\hexnumber\bmsfam}
  \mathchardef\bnabla="0\bsy@72
\fi
\makeatother
\title[]{Parsec-scale properties of a complete sample of 
radio galaxies}

\author[T. Venturi {\it et al.\/}]%
{T. V\ls E\ls N\ls T\ls U\ls R\ls I$^1$, 
\ns G.\ls G\ls I\ls O\ls V\ls A\ls N\ls N\ls I\ls N\ls I$^2$\ns\\
L.\ls F\ls E\ls R\ls E\ls T\ls T\ls I$^1$\ns 
W.\ls D.\ls C\ls O\ls T\ls T\ls O\ls N$^3$\ns 
\and \ns L.\ls L\ls A\ls R\ls A$^4$}

\affiliation{$^1$Istituto di Radioastronomia, CNR, Bologna, Italy\\
$^2$ Dipartimento di Fisica, Universit\`a di Bologna, Italy \\
$^3$ National Radio Astronomy Observatory, Charlottesville, VA, USA \\
$^4$ Instituto de Astrof\`sica de Andaluc\`ia (CSIC), Granada, Spain}

\begin{document}
\ifnfssone
\else
  \ifnfsstwo
  \else
    \ifoldfss
      \let\mathcal\cal
      \let\mathrm\rm
      \let\mathsf\sf
    \fi
  \fi
\fi

\maketitle

\begin{abstract}

We report the most important results on the parsec-scale
properties of a complete sample of radio galaxies carried out 
at radio frequencies 
with the Global VLBI array and with the VLBA.
Relativistic parsec-scale jets are common both
in FRI and FRII radio galaxies, and their orientation
to the line of sight is in agreement with the
expectations from the unified schemes for radio loud
AGNs. Proper motion has been detected in a few FRI galaxies
in the sample. FRI and FRII radio galaxies exhibit very
similar properties on the parsec-scale. Finally 
a few radio sources in the sample show evidence
of velocity structure in the parsec-scale jet, i.e. a
central spine with high Lorentz factor $\gamma$ and 
slower layers.

\end{abstract}

\firstsection 

\section{Introduction}

The study of the parsec-scale morphology and properties
of radio galaxies is essential to address a number
of issues. In particular it is of crucial importance 
{\it (a)} to understand the nature of the central engine 
and how radio  jets are born and propagate thoughout the
galaxy; and {\it (b)} to test the role of orientation and 
obscuration in the observed properties for this class of objects. 
It is now widely accepted (see for example Urry \& Padovani 1995) 
that the observed properties of AGNs depend mostly
on their orientation to the line of sight and on the
presence of a thick torus surrounding the central galactic
regions. According to this unified view of AGNs, FRI
and FRII radio sources would be the misaligned parent population
respectively of BL-Lacs and quasars.
Both classes of objects are therefore expected to 
be aligned at large angles to the line of sight and to be characterised
by relativistic speeds. Recent works of
Chiaberge et al. (1999 and 2000), based on observational data
in a wide range of frequencies, threw new insight on the
nuclear properties of FRIs and their relation to the
aligned parent population. In particular, they showed that
optical compact cores exist in FRIs, suggesting that obscuration
may not be relevant in these objects, and that thick tori are
not an essential component of their nuclear region. Furthermore,
they suggested that a two-phase jet in BL-Lacs and FRIs could
account for some discrepancies still defying unification.

\section{Parsec-scale imaging a complete sample of radio galaxies}

In the light of the results mentioned above, the knowledge
of the source orientation and plasma speed in the closest
vicinity of the nucleus is particularly important. This piece
of information is now available for a complete sample of
radio galaxies (Giovannini et al. 1990). The sample includes 27
galaxies, including 13 FRIs, 6 FRIIs, 2 BL-Lacs, 
1 compact symmetric object (CSO), one compact steep spectrum source 
(CSS) and four core/halo sources, with redshift ranging from
0.0021 to 0.162. All sources in the sample 
were observed and imaged at milliarcsecond 
resolution with the global VLBI array at 5 GHz during the past
decade. 
For a number of sources 1.6 GHz and 8.4 GHz observations
and multiepoch observations are also available.
The angular resolutions are of the order of $\theta_{FWHM} \sim 1 - 2$ mas,
corresponding to linear resolutions ranging from a few tenth of 
parsec to a few parsecs depending on the source distance.\\
Here we will concentrate on our results for FRI and FRII radio 
galaxies.

\section{Results and discussion}

\subsection{Radio morphologies}

Most FRI and FRII radio galaxies in the sample are characterised
by asymmetric morphology on the parsec-scale, with the mas jet
aligned with the dominant arcsecond scale one. In those cases
where multifrequency observations are available, the most compact
component turned out to be the core of the radio emission
(Cotton et al. 1999;
Giovannini et al. 1999a; 
Lara et al. 1997; 
Venturi et al. 1995;
Giovannini et al. 1994).
%

\noindent
We also found two symmetric sources, i.e. the FRI 3C338 (Giovannini
et al. 1998 and references therein) and the FRII 3C452 (Venturi et
al. 2000; Giovannini et al. in preparation). The latter is a narrow
line radio galaxy, and its two-sided parsec-scale morphology is
consistent with the idea that it is oriented at a large angle 
to the line of sight. For this source we derived $\theta \ge 60^{\circ}$.
Observational constraints for 3C338 suggest that this source 
lies almost in the plane of the sky, i.e. $\theta \sim 85^{\circ}$.
It is interesting to note that HST images for the host galaxy
show the presence of a compact nuclear component (Chiaberge et al. 1999),
therefore  obscuration is negligible for this object.

\noindent 
In Figure 1 we show the VLBI radio images of two typical radio
galaxies in the sample, i.e. the one-sided 3C66B and the symmetric
3C452.
Most parsec-scale jets in our images are similar to that in 3C66B,
where the jet brightness is centrally peaked.
In some cases the jets show high brigthness knots;
in other cases the jet brightness smoothly decreases
along the jet propagation direction (Venturi et al. 1995).

\begin{figure} 
   \vspace{2.5pc}
\centerline{\psfig{file=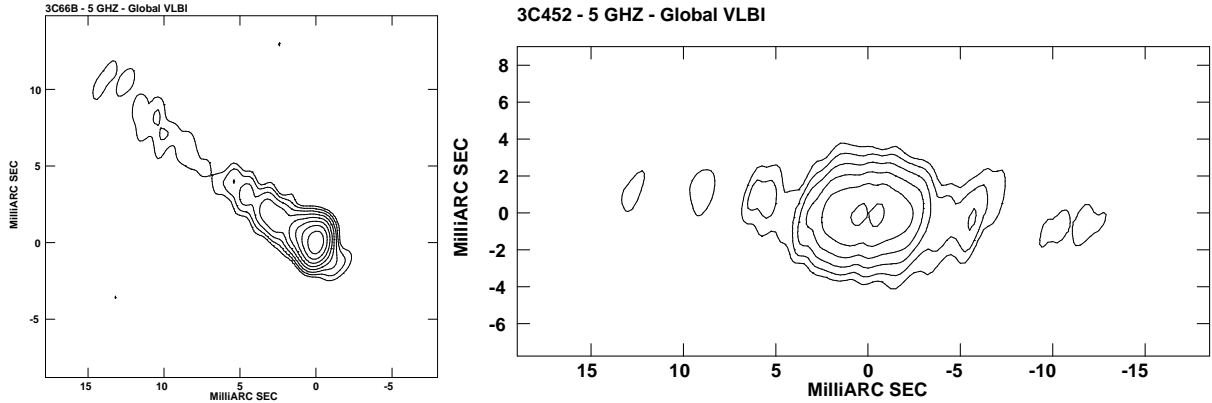,height=6.0truecm}}
  \caption{5 GHz global VLBI images of 3C66B (left) and 3C452 
(right). The restoring beam is respectively 
1.6$\times$0.8, p.a. $-15^{\circ}$ and
3.1$\times$0.7, p.a. $-13^{\circ}$. In both images the rms is $\sim 0.11$
mJy/beam and contours are -0.5, 0.5, 1, 2, 4, 8, 16 ... mJy/beam.
The linear scale is 1 mas = 0.6 pc and 2 pc for 3C66B and 3C452
respectively.}
\end{figure} 
There are however two remarkably different cases: for B2 1144+35,
shown in Figure 2, 
and MKN 501 (Giovannini et al. 1999a and 2000 respectively) 
the jets are limb-brightened. This could be explained assuming
that the jet consists of two different components, i.e. a fast central
spine,  deboosted if seen under viewing
angles $\theta < \gamma^{-1}$, and a surrounding shear-layer with
slower, but still relativistic, speed. This result is in 
agreement with the central spine-shear layer model predicted
by Laing (1996) for parsec-scale jets, and with the suggestion
made by Chiaberge et al. (2000).

\begin{figure} 
   \vspace{3.5pc}
\centerline{\psfig{file=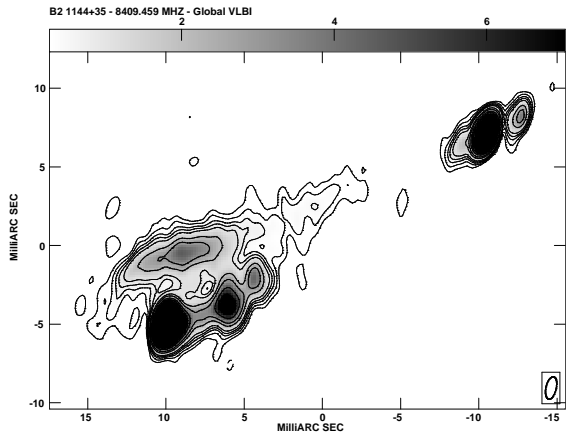,height=7.0truecm}}
  \caption{8.4 GHz global VLBI image of B2 1144+35.
The restoring beam is 1.45$\times$0.66, p.a. $-13^{\circ}$;
the rms is $\sim 0.05$
mJy/beam and contours are -0.2, 0.15, 0.2, 0.7, 0.7, 1, 2, 3, 5, 7,
10, 20, 30, 50 mJy/beam. Greyscale ranges from 0 to 7 mJy/beam.
The linear scale is 1 mas = 1.62 pc. }
\end{figure}

\subsection{Superluminal motion}
A major result found from our study is the detection of 
parsec-scale proper motion in FRI radio galaxies.
Beyond the well known case of the accelerating superluminal
jet in 3C274 (Biretta et al. 1999),
we found proper motion for three more FRIs in the sample,
i.e. NGC315 (Cotton et al. 1999), B2 1144+35 (Giovannini et
al. 1999a), 3C338 (Giovannini et al. 1998) and for the two 
BL-Lacs (Giovannini et al. 1999b, 2000). The proper
motion, $\beta_{app}$, found for the three FRIs 
ranges from $\sim 0.9$ (3C338) to 
2.7 (B2 1144+35). For the two BL-Lacs we found $\beta_{app}$ = 1.5
and 6.7 for MKN 421 and MKN 501 respectively.\\
It is noteworthy that there is evidence of an accelerating
parsec-jet in NGC315 (Cotton et al. 1999), where the intrinsic speed 
along the jet increases from $\beta_{intr} \sim$ 0.75 to 0.95
going from $\sim$ 3.4 to 9.5 mas from the core.

\subsection{Orientation and Lorentz factors}

Assuming that the observed morphologies and properties
of the galaxies in our sample are due to Doppler
boosting in an intrinsically symmetric source, we derived
possible ranges for the viewing angle $\theta$ and for  
the intrinsic speed $\beta_{intr}$ of the radio emitting plasma 
using standard beaming indicators (Giovannini et
al. 1994, and in preparation): 

\noindent
{\it (a)} the jet to counterjet
brightness ratio; 

\noindent
{\it (b)} the 5 GHz arcsecond scale core dominance 
with respect to the source total power at 408 MHz; 

\noindent
{\it (c)}
the superluminal motion, where available.

The viewing angles we derived both for FRIs and FRIIs are consistent
with the idea that both classes of sources are oriented at moderate
to large angles to the line of sight, i.e. $\theta \ge 30^{\circ}$,
in agreement with unification. The only exceptions to this rule
are B2 1144+35 and 3C274 (see Giovannini et al. 1999a and Biretta
et al 1999 for more details).
The intrinsic velocities found for FRIs are at least mildly
relativistic, with lower limits ranging from $\beta_{intr} \ge 0.4$
to 0.95.\\
Similar results are found for the FRIIs, reinforcing
the idea that FRIs and FRIIs are indistinguishable on the 
parsec-scale.\\
The limits to the Lorentz factors we derived are 
in the range 3 -10, 
in agreement with the results
obtained by Chiaberge et al. (2000), who require $\gamma$ in the
range 3 - 7 to account for the different orientation properties
in FRIs and BL-Lacs.


\end{document}